\documentclass[reprint,
preprintnumbers,
amsmath,amssymb,
aps,prd
]{revtex4-2}

\usepackage{tikz}
\usetikzlibrary{arrows.meta, positioning, calc}

\usepackage{graphicx}
\usepackage{dcolumn}
\usepackage{bm}
\usepackage{xcolor}
\RequirePackage{xspace}

\usepackage{listings}
\usepackage{xcolor}

\lstdefinelanguage{Julia}{
  morekeywords={
    abstract,break,case,catch,const,continue,do,else,elseif,end,export,false,finally,
    for,function,global,if,import,in,let,local,macro,module,mutable,primitive,quote,
    return,struct,try,true,using,while
  },
  sensitive=true,
  morecomment=[l]\#,
  morestring=[b]",
}

\newcommand{\im}{\mathrm{i}}
\newcommand{\sun}{$\mathop{\rm SU}(N_c)$\xspace}
\newcommand{\suthree}{$\mathop{\rm SU}(3)$\xspace}

\begin{document}

\preprint{UTHEP-821, UTCCS-P-178}

\title{Lattice Gauge Theory via LLVM-Level Automatic Differentiation}

\author{Yuki Nagai}
\email{nagai.yuki@mail.u-tokyo.ac.jp}
\affiliation{
Information Technology Center, The University of Tokyo,
6--2--3 Kashiwanoha, Kashiwa, Chiba 277--0882, Japan
}
\affiliation{
Department of Advanced Materials Science, The University of Tokyo, Kashiwa, Chiba 277-8561, Japan
}

\author{Hiroshi Ohno}
\affiliation{
Center for Computational Sciences, University of Tsukuba,
1-1-1 Tennodai, Tsukuba, Ibaraki 305-8577, Japan
}
\email{hohno@ccs.tsukuba.ac.jp}

\author{Akio Tomiya}
\affiliation{
Department of Information and Mathematical Sciences, Division of Mathematical Sciences,
Tokyo Women's Christian University,
2-6-1 Zempukuji, Suginami-ku, Tokyo 167-8585, Japan
}
\affiliation{RIKEN Center for Computational Science, Kobe 650-0047, Japan}
\affiliation{
Department of Physics, Kyoto University, Kyoto 606-8502, Japan
}
\email{akio@yukawa.kyoto-u.ac.jp}

\date{\today}

\begin{abstract}
We enable the automatic construction of Hybrid Monte Carlo (HMC) forces in lattice gauge theory by performing reverse-mode automatic differentiation at the level of optimized LLVM intermediate representation, making the approach applicable to any language that lowers lattice action code to LLVM.
In practice, this means that once the action evaluation routine is implemented, the corresponding HMC force can be generated automatically from the same code path, without deriving or maintaining a separate force routine.
The method preserves conventional imperative, in-place implementations and enables a single-source workflow in which forces are generated directly from the action code while inheriting compiler optimizations. We perform end-to-end reverse-mode differentiation of both gauge and Wilson fermion actions. 
For the Wilson fermion case, we find that the force generated by automatic differentiation achieves performance comparable to a conventional hand-written fermion force implementation.
The same differentiation pipeline targets both CPU and GPU backends, providing a practical route to performance-portable force construction for compositional lattice actions.
\end{abstract}

\maketitle

\section{Introduction}
Lattice gauge theory provides a systematically improvable, first-principles approach to strongly coupled gauge theories, including quantum chromodynamics (QCD).
Even in regimes where perturbation theory breaks down, lattice calculations enable controlled determinations of hadron spectra~\cite{CP-PACS:2001vqx}, QCD matrix elements~\cite{Tsang:2025gdc}, and thermodynamic observables~\cite{Schmidt:2025ppy}, and thereby support precision Standard Model baselines relevant to searches for physics beyond the Standard Model~\cite{DelDebbio:2018szp}.
Realistic simulations with dynamical fermions rely on the Hybrid Monte Carlo (HMC) algorithm~\cite{Duane:1987de}, which is the de facto standard:  an auxiliary Hamiltonian system is evolved by molecular dynamics (MD) and corrected by a Metropolis accept--reject step.

A central computational object in HMC is the MD force, given by the variation of the lattice action with respect to the link variables $U_\mu(x)$.
In modern lattice workflows, however, the action used in production is rarely a simple textbook form.
Improved gauge actions, rational approximations in Rational HMC (RHMC), and multi-level link smearing (e.g.\ stout smearing \cite{Morningstar:2003gk}) introduce a layered composition of transformations that deepens both the algebraic structure and the implementation complexity~\cite{Clark:2006wq}.
In many established code bases, one must (i) derive the force analytically for each action and (ii) implement a separate force routine whose correctness is then validated against the action evaluation.
As actions become more compositional, these two steps increasingly dominate development and validation cost, and force/action inconsistencies become a frequent source of subtle bugs.

At the same time, lattice simulations are migrating toward heterogeneous high-performance computing (HPC) environments centered on accelerators \cite{Boyle:2022ncb}.
Effective use of GPUs is now practically mandatory on flagship systems, while GPU execution models and performance characteristics vary across vendors and generations.
From the standpoint of long-term software sustainability, it is desirable to reduce the coupling between the {physics-side} definition of the action and the {architecture-specific} implementation of the force, without abandoning in-place updates and explicit control flow that are standard in performance-oriented lattice codes.

Automatic differentiation (AD), also called algorithmic differentiation, is now ubiquitous in machine learning because it provides program-level derivatives with little manual algebra and supports efficient reverse-mode backpropagation, and this differentiable-programming ecosystem is increasingly used for learned-proposal gauge-configuration generation, including normalizing-flow samplers and flow-augmented Hybrid Monte Carlo (HMC) \cite{Tomiya:2025quf,Cranmer:2023xbe,Albergo:2019eim,Foreman:2021ljl,Foreman:2023ymy}. In our study of gauge-covariant network \cite{Nagai:2021bhh}, we pointed out that differentiating with respect to gauge links extends the reverse-mode ``delta rule'' to gauge-group variables, so that treating a gauge flow as a neural network yields well-defined link gradients while preserving gauge covariance \cite{Nagai:2021bhh,Nagai:2025rok}. Building on this viewpoint, gauge equivariant/covariant flows on \sun have been developed and generalized, extended to fermionic theories, and connected to a variety of proposal mechanisms, alongside complementary efforts that learn trivializing flows and related maps \cite{Kanwar:2020xzo,Boyda:2020hsi,Abbott:2023thq,Komijani:2025yjz,Albergo:2021bna,DelDebbio:2021qwf,Bacchio:2022vje,Albandea:2023wgd}. Nonequilibrium reweighting and stochastic-flow protocols, together with differentiable effective-action and RG-inspired constructions, further broaden the design space for gauge-configuration generation \cite{Caselle:2016wsw,Caselle:2022acb,Caselle:2022esc,Bulgarelli:2024brv,Bonanno:2023ier,Holland:2024muu,Wenger:2025sre}.

AD is a natural candidate for constructing forces from action programs.
However, AD in general purpose machine learning frameworks is optimized for tensor-centric, functional computation graphs and often mismatches lattice codes that rely on explicit loops, branching, and aggressive in-place reuse of temporaries 
\cite{abadi2016tensorflowlargescalemachinelearning, paszke2019pytorchimperativestylehighperformance}.
Moreover, tape-based reverse-mode AD may require storing intermediate states whose memory footprint is prohibitive for long HMC trajectories on large lattices.
These issues motivate an alternative that is closer to the compiler representation used to optimize and lower HPC kernels \cite{moses2020instead}.

In this work we construct HMC forces by differentiating lattice action codes at the level of LLVM intermediate representation (LLVM IR) \cite{lattner2004llvm} using an LLVM-level reverse-mode AD system \cite{moses2020instead} to generate the adjoint program corresponding to the optimized action-evaluation routine.
Our implementation is written in Julia~\cite{bezanson2012juliafastdynamiclanguage}, which uses LLVM-based just-in-time compilation and is known to achieve performance comparable to C and Fortran for numerical workloads \cite{bezanson2012juliafastdynamiclanguage,regier2018catalogingvisibleuniversebayesian}; however, Enzyme operates directly on LLVM IR and is therefore applicable to any frontend that lowers to LLVM, including Fortran, C++ and Rust.
The resulting force routine is obtained directly from the action code, preserving an imperative, in-place style and inheriting compiler optimizations.
We demonstrate compiler-generated construction of HMC forces for both gauge and Wilson fermion actions, including multi-level stout smearing, via LLVM-level reverse-mode differentiation, targeting both CPU and GPU backends through JACC.jl~\cite{JACC}.
We validate the generated force for the Wilson fermion action at the link level against a conventional hand-written implementation and confirm dynamical consistency in full HMC trajectories.

\begin{figure}[t]
\centering
\begin{tikzpicture}[ 
  font=\scriptsize,
  box/.style={draw, rounded corners, align=center, inner xsep=4pt, inner ysep=3pt},
  arr/.style={-Latex, line width=0.5pt},
  lab/.style={align=center, font=\scriptsize},
  note/.style={align=left, font=\scriptsize}
]
\node[lab] (fwtitle) {\textbf{Forward: SSA values along the optimized instruction sequence}};
\node[box, below=2mm of fwtitle] (U0) {\texttt{\%U0}\\$U_0\in $ \sun};
\node[box, below=7mm of U0] (U1) {\texttt{\%U1}\\$U_1$};
\node[box, below=7mm of U1] (U2) {\texttt{\%U2}\\$U_2$};
\node[box, below=7mm of U2] (UN) {\texttt{\%UN}\\$U_N$};
\node[box, below=7mm of UN] (S) {$S$\\(scalar)};

\draw[arr] (U0) -- (U1) node[midway, right=2mm, lab]{smear};
\draw[arr] (U1) -- (U2) node[midway, right=2mm, lab]{smear};
\draw[arr] (U2) -- (UN) node[midway, right=2mm, lab]{Dirac/CG\\+ contractions};
\draw[arr] (UN) -- (S)  node[midway, right=2mm, lab]{accumulate};

\node[note, right=0mm of U1, anchor=west] (ssanote) {%
\textbf{SSA viewpoint}\\
in-place code $\Rightarrow$ distinct values \texttt{\%Ui}\\
(memory reuse hidden)
};

\node[box, right=15mm of U2, anchor=west, align=left] (llvm) {%
\textbf{LLVM-like trace}\\
\texttt{\%U1 = stout(\%U0)}\\
\texttt{\%U2 = stout(\%U1)}\\
\texttt{\%S  = action(\%UN)}
};

\node[lab, below=8mm of S] (rvtitle) {\textbf{Reverse: adjoint traversal of the same instructions}};
\node[box, below=2mm of rvtitle] (Sbar) {$\bar S=1$};
\node[box, below=7mm of Sbar] (UbarN) {\texttt{\%$\bar U$N}\\$\bar U_N$};
\node[box, below=7mm of UbarN] (Ubar2) {\texttt{\%$\bar U$2}\\$\bar U_2$};
\node[box, below=7mm of Ubar2] (Ubar1) {\texttt{\%$\bar U$1}\\$\bar U_1$};
\node[box, below=7mm of Ubar1] (Ubar0) {\texttt{\%$\bar U$0}\\$\bar U_0$};

\draw[arr] (Sbar) -- (UbarN) node[midway, right=2mm, lab]{seed};
\draw[arr] (UbarN) -- (Ubar2) node[midway, right=2mm, lab]{reverse};
\draw[arr] (Ubar2) -- (Ubar1) node[midway, right=2mm, lab]{reverse};
\draw[arr] (Ubar1) -- (Ubar0) node[midway, right=2mm, lab]{reverse};

\node[box, right=8mm of Ubar0, anchor=west] (force) {read out\\HMC force};
\draw[arr] (Ubar0.east) -- (force.west);
\end{tikzpicture}
\caption{
Compiler-level view of HMC force construction as the adjoint of an in-place action evaluation.
In LLVM SSA form, apparent in-place updates correspond to a sequence of distinct values \texttt{\%Ui}.
Enzyme generates the reverse pass by traversing the optimized instruction sequence backward and propagating adjoints \texttt{\%$\bar U$i}.
The force is obtained from the adjoint associated with the initial SSA value (followed by Lie-algebra projection).
}
\label{fig:ssa_adjoint}
\end{figure}

\section{\label{sec:LLVM}LLVM-level AD for differentiable lattice gauge theory}

\subsection{Action evaluation as discrete-time program dynamics}
In lattice simulations, an ``action'' is not manipulated as a symbolic functional but evaluated as a concrete numerical program acting on gauge fields and auxiliary variables.
Even for an action formally written as $S[U]$, its numerical evaluation proceeds through an ordered sequence of operations: link smearing, construction of derived fields, application of Dirac operators, linear solves, and contractions to a scalar.
This motivates a viewpoint in which action evaluation is a discrete-time evolution of \emph{program states},
\begin{align}
U_0 \;\longrightarrow\; U_1 \;\longrightarrow\; \cdots \;\longrightarrow\; U_N \;\longrightarrow\; S,
\label{eq:program_flow}
\end{align}
where $U_0$ is the original gauge field, $U_i$ denotes intermediate source code-level objects (e.g.\ smeared links or temporary fields), and $S$ is the final scalar.
The ``time'' index is not physical time but the position along the instruction stream.
This perspective is independent of the specific discretization and naturally accommodates compositional actions, including deep smearing chains and fermion actions with nested solvers.

Crucially, lattice implementations typically use in-place updates for performance and memory efficiency.
At source level, this may look like overwriting a field, but algorithmically it represents a deterministic transition from one intermediate state to the next in Eq.~\eqref{eq:program_flow}.
To define a reverse pass without ambiguity, we need a representation in which these successive states are distinguished.
This is provided by static single assignment (SSA) form at the compiler IR level.

\subsection{SSA form and well-posed adjoint evolution}
LLVM IR is expressed in static single assignment (SSA) form: each intermediate value is defined exactly once.
As a result, an in-place update in source code corresponds, at IR level, to producing a new SSA value.
Successive ``updates'' of what appears to be the same object are represented as distinct SSA values
\begin{align}
\%U_0,\ \%U_1,\ \ldots,\ \%U_N,
\end{align}
each tied to a specific instruction (or basic-block output) in the optimized program.
This compiler-level viewpoint is schematically illustrated in Fig.~\ref{fig:ssa_adjoint}. 
The reuse of physical memory locations is an implementation detail handled by the compiler and backend; it does not erase the logical dependence structure encoded in SSA.
Therefore, the action evaluation in SSA form provides a well-defined discrete-time dynamical system in program space, together with a clear dataflow graph even when the source code is imperative and in-place.

This property is particularly important for lattice codes, where explicit load/store operations, stencil-like updates, halo exchanges, and carefully managed temporaries are standard.
Differentiating at the IR level means differentiating \emph{the lowered, optimized program} that the compiler actually executes.
The reverse pass is thus naturally aligned with the optimized forward computation, and the differentiation ``sees'' the same control flow and memory behavior that matters for performance.

\subsection{Adjoint evolution and the HMC force}

Given the SSA-level discrete-time evolution in Eq.~\eqref{eq:program_flow},
the HMC force is obtained by propagating sensitivities from the final
scalar back to the initial links.
For a schematic forward update
\begin{equation}
U_{i+1}=\mathcal{F}_i(U_i),
\end{equation}
the corresponding reverse (adjoint) update reads
\begin{equation}
\bar U_i=
\left(\frac{\partial \mathcal{F}_i}{\partial U_i}\right)^\dagger \bar U_{i+1},
\qquad
\bar U_i \equiv \frac{\partial S}{\partial U_i}.
\label{eq:adjoint_update}
\end{equation}
Although Eq.~\eqref{eq:adjoint_update} resembles the chain rule,
the essential point is methodological:
$\mathcal{F}_i$ represents a concrete program transformation
encoded in the optimized instruction stream.
In our framework, the adjoint evolution is generated
by differentiating that instruction sequence directly,
rather than by analytic manipulation of action formulas.

In the molecular-dynamics evolution we adopt the standard
left-multiplication update
\begin{equation}
U_\mu(x) \;\rightarrow\;
e^{\,\im\epsilon H_\mu(x)}\, U_\mu(x),
\qquad
H_\mu(x) \in \mathfrak{su}(N_c),
\label{eq:left_update}
\end{equation}
so that
$\delta U_\mu(x)= \im\,H_\mu(x)\,U_\mu(x)$.

Using the conventional matrix-derivative relation
\begin{equation}
\delta S =
\mathrm{Re}\,
\mathrm{tr}
\!\left[
\left(
\frac{\partial S}{\partial U_\mu(x)}
\right)^\dagger
\delta U_\mu(x)
\right],
\end{equation}
one finds that the Lie-algebra force entering the momentum update
is obtained by projection onto $\mathfrak{su}(N_c)$,
\begin{equation}
F_\mu(x)
=
\mathcal{P}_{\mathfrak{su}(N_c)}
\!\left[
\left(
\frac{\partial S}{\partial U_\mu(x)}
\right)
U_\mu^\dagger(x)
\right],
\label{eq:su_projection}
\end{equation}
where
$\mathcal{P}_{\mathfrak{su}(N_c)}[A]
=
\frac{1}{2}(A-A^\dagger)
-\frac{1}{2N_c}
\mathrm{tr}(A-A^\dagger)\mathbf{1}$.
Thus the compiler-generated adjoint provides the matrix derivative
$\partial S/\partial U$, from which the conventional HMC force
is constructed by right multiplication with $U^\dagger$
followed by Lie-algebra projection.
Details of the matrix-derivative convention,
its relation to the real-component differentiation performed by Enzyme
(via Wirtinger calculus), and the construction of the Wilson fermion force
are provided in the Supplemental Material
(Secs.~S2--S3).

\subsection{Implementation via Enzyme at the LLVM level}
We realize the above construction using Enzyme~\cite{moses2020instead}, which performs reverse-mode AD directly on LLVM IR.
Given a Julia function that evaluates the action (action value and, where needed, intermediate contractions), Julia lowers the code to LLVM IR and applies standard compiler optimizations.
Enzyme then analyzes the optimized SSA instruction stream, determines the required data dependencies for the reverse pass, and generates an adjoint routine that propagates adjoint values in reverse execution order.
Because differentiation happens at IR level, Enzyme naturally handles explicit load/store operations and in-place updates as they appear after lowering.
It also differentiates complex arithmetic as implemented at the LLVM level, avoiding manual splitting into real and imaginary parts.

From the user perspective, the workflow is: write an action evaluation routine in a conventional imperative style (including explicit loops and in-place updates); apply Enzyme to obtain the corresponding adjoint routine; post-process by Lie-algebra projection to obtain the HMC force.
Importantly, this is neither source-to-source symbolic differentiation nor a tape-based recording of runtime operations.
It is a compiler-level transformation of the optimized program, and thus inherits the optimization structure that is essential for HPC.
While reverse-mode differentiation may still require intermediate values, this requirement is determined by Enzyme's analysis of the optimized IR, and it can exploit compiler-level simplifications.
In the present work we do not employ additional checkpointing; instead we focus on demonstrating end-to-end correctness and practical viability on current accelerator hardware.

Our implementation is built on the JuliaQCD ecosystem.
Gauge fields and link-smearing operations are provided by \texttt{Gaugefields.jl} and \texttt{LatticeMatrices.jl}, while fermion operators and fermion actions are constructed using \texttt{LatticeDiracOperators.jl}~\cite{Nagai:2024yaf}.
This modular structure allows actions built from these components (including deep stout-smearing chains) to be expressed in a single code path that is differentiable at LLVM level.

\section{\label{sec:HPC}In-place HPC implementation, validation, and portability}

A practical force evaluation must be compatible with in-place execution, reuse of temporaries, and accelerator backends.
Our implementation follows the standard lattice-code approach of explicit loops and in-place updates, and differentiates the resulting optimized LLVM IR using Enzyme.
GPU execution is enabled via the JACC.jl backend~\cite{JACC}; in this setup the same Julia source code and the same Enzyme-based differentiation pipeline can be used on both CPUs and GPUs.
This provides a path to performance-portable force construction while keeping the physics-side action definition in a single code path.

In the current JACC.jl-based GPU backend, reverse-mode differentiation through certain low-level device kernels is not directly supported. In particular, site-local matrix multiplications defined on the lattice and offloaded via JACC require explicitly defined adjoints to ensure correct device execution. 
We therefore provide custom reverse implementations for these selected operations. These adjustments reflect practical integration constraints between LLVM-level automatic differentiation and the GPU backend, but they do not modify the conceptual framework in which the HMC force is obtained as the adjoint of the action-evaluation routine.

\subsection{Local validation: link-level force agreement for the Wilson fermion action}
We first validate the force generated by LLVM-level AD for the Wilson fermion action, for which a conventional force construction is available and widely used in dynamical-fermion HMC implementations. The automatically generated force, $F_{\rm AD}$, is compared with a reference force evaluated by the standard approach, $F_{\rm ana}$, at the level of individual link matrices across the lattice. The comparison is performed for \suthree gauge theory on a $16^4$ lattice with the Wilson–Dirac operator.

\begin{figure}[h]
\begin{center}
\includegraphics[width=0.45\textwidth]{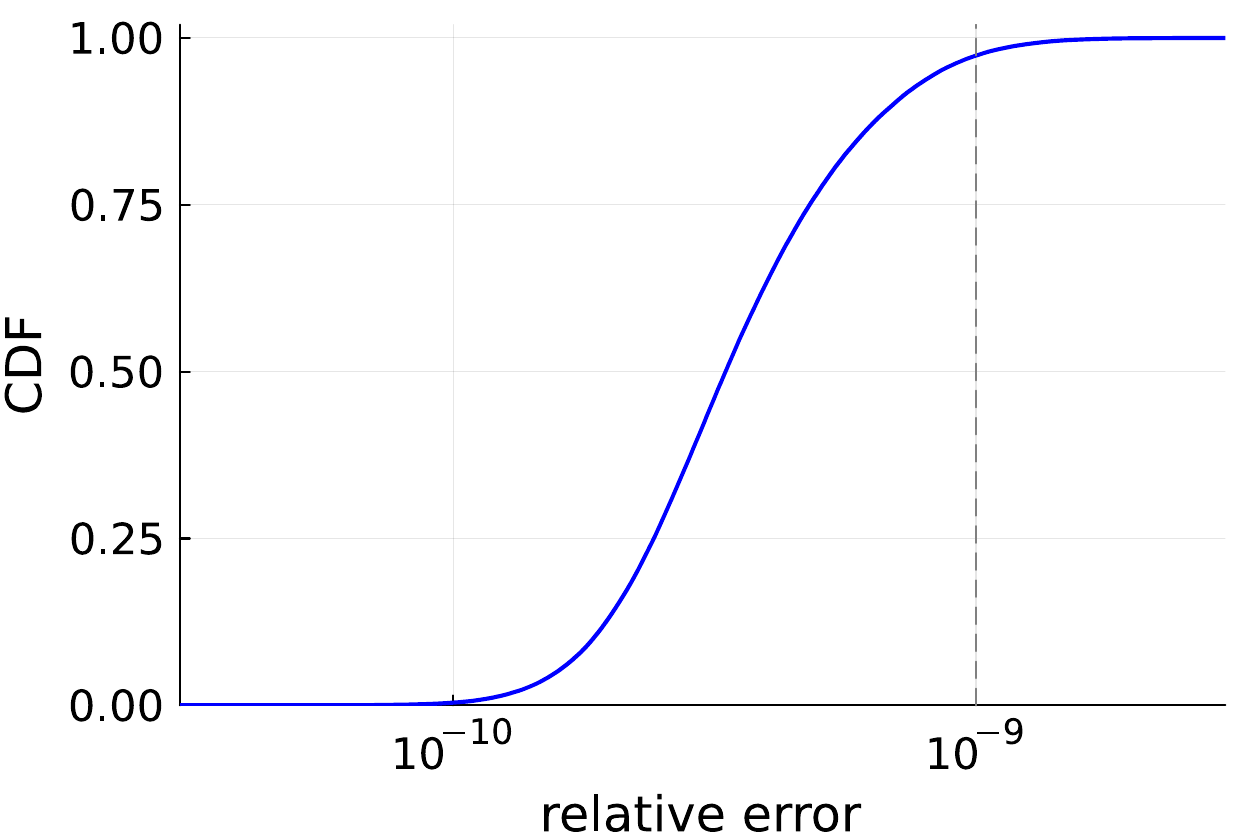}
\end{center}
\caption{Cumulative distribution of the relative Frobenius-norm error between the molecular-dynamics force for the Wilson fermion action generated by LLVM-level automatic differentiation, $F_{\rm AD}$, and the corresponding force computed by a conventional hand-written implementation, $F_{\rm ana}$, on an \suthree $16^4$ lattice. The error is evaluated link by link over the lattice. Agreement is observed at the $10^{-9}$ level, consistent with double-precision roundoff.}
\label{fig:cdf}
\end{figure}

Figure~\ref{fig:cdf} shows the cumulative distribution of the relative Frobenius-norm error,
\begin{align}
\epsilon(x,\mu) \equiv
\frac{\|F_{\rm AD}(x,\mu)-F_{\rm ana}(x,\mu)\|_{\rm F}}{\|F_{\rm ana}(x,\mu)\|_{\rm F}},
\label{eq:force_error}
\end{align}
with $\|A\|_{\rm F}\equiv \sqrt{\mathrm{tr}(A^\dagger A)}$, 
evaluated over all links. We observe uniform agreement at the level of $10^{-9}$, consistent with double-precision roundoff, with no visible outliers. This establishes that the compiler-generated adjoint reproduces the reference Wilson fermion force at the link level and validates the end-to-end differentiation pipeline in a dynamical-fermion setting.

\subsection{Dynamical validation: energy conservation and step-size scaling}
Local force agreement is necessary but not sufficient for stable HMC evolution.
A more stringent test is to examine the energy violation $\Delta H \equiv H_{\mathrm{final}} - H_{\mathrm{initial}}$ along complete HMC trajectories.
For a reversible and symplectic integrator of second order (e.g.\ leapfrog), the expectation value scales as $\langle \Delta H \rangle \propto \Delta t^2$, 
where $\Delta t$ is the MD step size.
This scaling is sensitive to inconsistencies between the action and the implemented force and therefore serves as a dynamical validation of the full pipeline.

\begin{figure}[h]
\begin{center}
\includegraphics[width=0.5\textwidth]{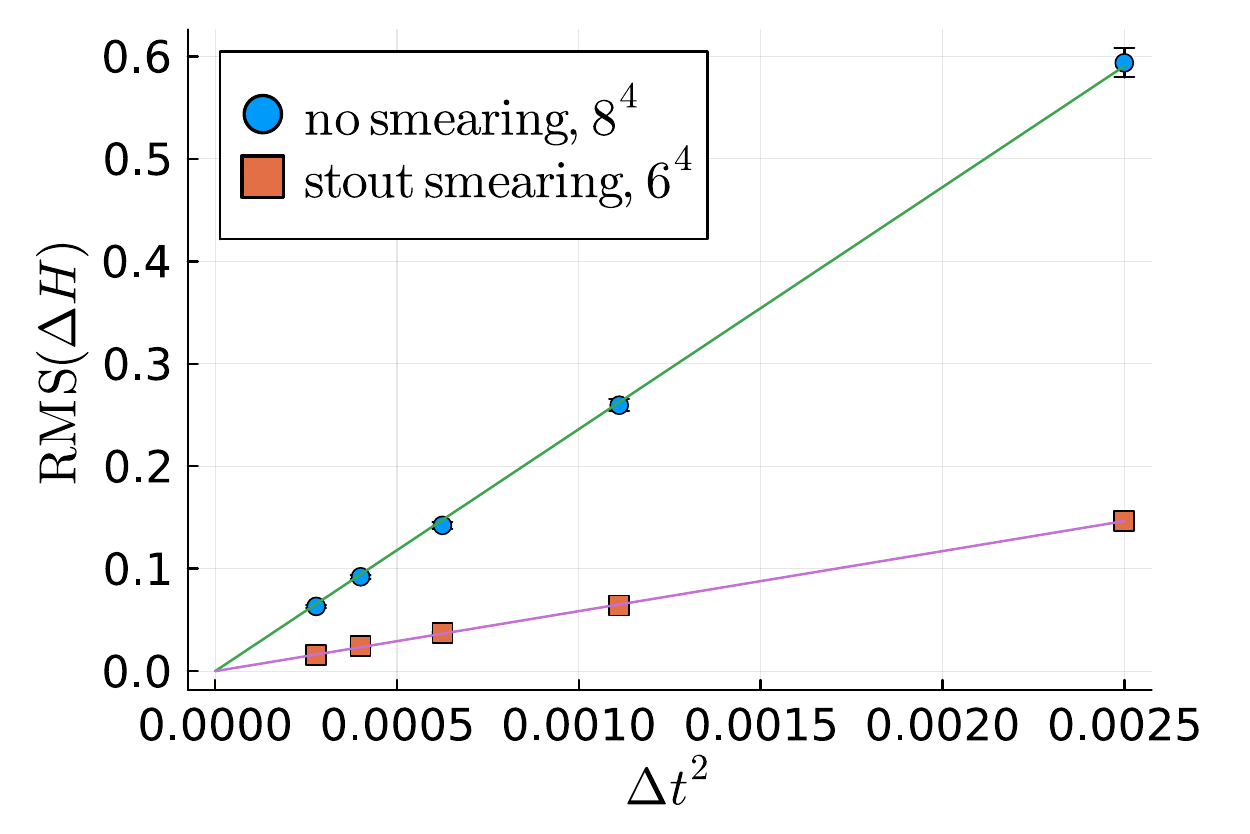}
\end{center}
\caption{
Step-size dependence of HMC diagnostics using automatically generated forces for Wilson fermion simulations. The data exhibit the expected $\Delta H \propto \Delta t^2$ scaling characteristic of a second-order symplectic integrator, together with stable acceptance behavior. Blue circles denote standard Wilson fermion HMC on an $8^4$ lattice without stout smearing. Red squares correspond to simulations on a $6^4$ lattice with stout smearing ($\rho=0.3$), combined with a Sexton–Weingarten multiple time-scale integrator with 10 sub-steps.
}
\label{fig:b}
\end{figure}

We perform HMC simulations for the Wilson gauge action on an $8^4$ lattice without smearing using the automatically generated force. For comparison, we also consider simulations on a $6^4$ lattice with stout smearing ($\rho=0.3$), combined with a Sexton–Weingarten multiple time-scale integrator employing 10 inner steps. The first 100 trajectories are discarded for thermalization, and measurements are accumulated over the remaining trajectories. 
Figure~\ref{fig:b} shows the step-size dependence of $\Delta H$ for various choices of the number of MD steps. In both cases, the data exhibit the expected quadratic scaling behavior, with no sign of systematic drift, demonstrating full consistency between the forward action and the compiler-generated adjoint routine.

Unlike the analytically simple Wilson-gauge case, the present setup involves Wilson fermions with stout smearing ($\rho=0.3$) and a Sexton–Weingarten multiple time-scale integrator. Although only a single level of smearing is employed, the force structure is already considerably more involved than in the pure gauge case. The observed stability of HMC and the expected quadratic scaling of $\Delta H$ demonstrate that the compiler-generated adjoint faithfully reproduces the required force contributions in this nontrivial setting.

\subsection{Performance characteristics}

We summarize representative timings for the Wilson fermion action without smearing.
The measurements correspond to a single force evaluation (one MD step) on a $24^4$ lattice.
CPU results are obtained in single-threaded mode, and GPU results are obtained on an NVIDIA H100.
Table~\ref{tab:timing} compares two implementations based on the \texttt{LatticeMatrices.jl} backend: a manually implemented force and an LLVM-AD generated force.

\begin{table}[t]
\caption{
Representative wall-clock time (seconds) for a single force evaluation
of the Wilson fermion action on a $24^4$ lattice.
CPU results correspond to single-thread execution.
GPU results are obtained on an NVIDIA H100 using JACC.jl.
}
\begin{ruledtabular}
\begin{tabular}{lcc}
Implementation & CPU & GPU \\
\hline
Hand-written & 195.53 & 4.61 \\
LLVM-AD & 162.50 & 3.39 \\
\end{tabular}
\end{ruledtabular}
\label{tab:timing}
\end{table}

Within the same backend framework, LLVM-level AD produces a force routine that is competitive with a manually implemented version on both CPU and GPU.
On CPU, the LLVM-AD force is moderately faster than the corresponding hand-written implementation in this setup.
On GPU, the LLVM-AD force is likewise slightly faster than the hand-written force within the same backend.
These results indicate that IR-level differentiation does not introduce significant overhead and can achieve performance comparable to carefully implemented analytic force routines when built on the same matrix backend.

Absolute performance depends on backend maturity and compiler optimizations.
The present comparison is therefore intended to demonstrate parity within a common implementation framework rather than to establish absolute optimal performance.

The memory footprint for the $24^4$ lattice on the H100 was approximately 22\,GB in the present setup.
No special memory-reduction techniques (such as explicit checkpointing) were applied.
This indicates that the LLVM-level adjoint construction can be used in practice on current accelerator hardware for representative lattice sizes, while leaving room for future improvements in memory reduction and performance tuning.

What we demonstrate here is that HMC forces can be generated directly from optimized action code at the LLVM level.
Once an action evaluation is implemented in a conventional imperative, in-place style, the corresponding HMC force is obtained automatically from the same code path, without analytic derivation and without maintaining a separate force kernel.
This reduces development effort, lowers the risk of inconsistencies between action and force, and enables rapid exploration of improved actions and smearing chains.
Combined with GPU portability through JACC.jl, it offers a practical route to single-source, performance-portable force construction for compositional lattice actions.

\section{Summary and outlook\label{sec:Summary}}
We presented a compiler-level framework for constructing HMC forces by differentiating lattice action programs at the level of LLVM intermediate representation.
The key conceptual step is to view action evaluation as a discrete-time evolution of program states and to exploit SSA form to render in-place updates as a sequence of distinct intermediate values.
In this representation, the HMC force is obtained as the adjoint traversal of the optimized instruction stream.
Using Enzyme integrated in Julia, we performed end-to-end differentiation of gauge and fermion actions, validated the resulting force against the analytic Wilson fermion force at the link level, and confirmed dynamical consistency through HMC energy-violation diagnostics, including the expected $\Delta H \propto \Delta t^2$ scaling.

Representative performance measurements indicate that LLVM-level AD can generate forces that are competitive with hand-written implementations within the same backend, on both CPU and GPU.
While absolute peak performance depends on backend optimization and may still favor highly tuned, architecture-specific kernels, the compiler-level approach substantially reduces the combined development and maintenance cost by eliminating separate analytic force derivations and dedicated force routines.
This is particularly relevant for actions with deep compositional structure, such as multi-level smearing or algorithmically defined transformations.

Several directions are natural for future work.
First, memory reduction techniques (e.g.\ checkpointing strategies adapted to the compiler-generated adjoint) could extend applicability to larger volumes and longer trajectories.
Second, further optimization of the underlying backends and kernel fusion strategies may improve absolute performance.
HMC with AD may also be beneficial for future AI-driven automation in lattice gauge theory research.
Finally, while we demonstrated \sun gauge theory with Wilson fermions as a concrete case, the approach is not tied to a particular discretization as long as the action is expressed as a differentiable program; extending systematic benchmarks to other fermion formulations and gauge groups is an important next step.
Overall, 
LLVM-level automatic differentiation connects lattice gauge theory codes with modern differentiable-programming infrastructures in a concrete and implementation-level manner.
The implementation used in this work is available as part of the open-source JuliaQCD ecosystem. In particular, the LLVM-level automatic differentiation framework described here is included in the current release of \texttt{LatticeDiracOperators.jl}. Instructions for reproducing the AD-based force construction are provided in the repository README.

\begin{acknowledgments}
The work of A.T.\ was partially supported by JSPS KAKENHI Grant Numbers 20K14479, 22H05111, 22K03539 and 25K07287 and JST BOOST, Japan Grant Number JPMJBY24F1.
The work of A.T and H.O.\ was partially supported by JSPS KAKENHI Grant Number 22H05112.
The work of Y.N.\ was partially supported by JSPS KAKENHI Grant Numbers 22K03539, 22H05114  and 22K12052. 
This work was partially supported by "Joint Usage/Research Center for Interdisciplinary Large-scale Information Infrastructures (JHPCN)" in Japan (Project ID: jh250046).
This work was partially supported by MEXT as ``Program for Promoting Researches on the Supercomputer Fugaku''
(Grant Number JPMXP1020230411, JPMXP1020230409).
\end{acknowledgments}

\bibliography{apssamp}

\clearpage
\onecolumngrid

\setcounter{section}{0}
\setcounter{equation}{0}
\setcounter{figure}{0}
\setcounter{table}{0}

\renewcommand{\thesection}{S\arabic{section}}
\renewcommand{\theequation}{S\arabic{section}.\arabic{equation}}
\renewcommand{\thefigure}{S\arabic{figure}}
\renewcommand{\thetable}{S\arabic{table}}

\makeatletter
\@addtoreset{equation}{section}
\makeatother

\begin{center}
{\Large \bf Supplemental Material}\\[1em]
{\large Lattice Gauge Theory via LLVM-Level Automatic Differentiation}
\end{center}

\section{Overview}

This Supplemental Material provides implementation details
omitted from the main text. In particular, we clarify the relation
between LLVM-level automatic differentiation as implemented in
\texttt{Enzyme.jl} and the Wirtinger-derivative formulation
of variations on \sun link variables.
\section{Left-Invariant HMC, Matrix Derivatives, and the Enzyme Output}
\label{sec:SM-leftinv-enzyme}

This section clarifies how the LLVM-level derivative returned by
\texttt{Enzyme.jl} is related to the conventional matrix derivative
used in lattice gauge theory, and why our implementation constructs
the combination $U\,G^\dagger$ when building the HMC force.

\subsection*{Left-invariant HMC equations}

We adopt the left-multiplication form of the molecular-dynamics (MD)
evolution for link variables $U\in$ \sun and conjugate momenta
$P\in\mathfrak{su}(N_c)$,
\begin{align}
U(\tau+\Delta\tau)
&=
e^{\,\im\,\Delta\tau\,P(\tau)}\,U(\tau),
\label{eq:Uupdate-left}\\
P(\tau+\Delta\tau)
&=
P(\tau)-\Delta\tau\,\mathcal F(\tau),
\label{eq:Pupdate-left}
\end{align}
where $\mathcal F(\tau)\in\mathfrak{su}(N)$ denotes the HMC force.
The infinitesimal left variation
\begin{equation}
U \;\longrightarrow\; e^{\im\epsilon H}U,
\qquad H\in\mathfrak{su}(N_c),
\end{equation}
implies
\begin{equation}
\delta U = \im H U.
\label{eq:deltaU_left}
\end{equation}

\subsection*{Matrix derivative convention and force}

For a real-valued action $S(U)$ we use the standard lattice
matrix-derivative convention
\begin{equation}
\left[
\frac{\partial S}{\partial U}
\right]_{ij}
\equiv
\frac{\partial S}{\partial U_{ji}},
\label{eq:matrix_deriv_conv}
\end{equation}
so that first-order variations can be written as
\begin{equation}
\delta S
=
\mathrm{Re}\,
\mathrm{Tr}
\!\left[
\left(
\frac{\partial S}{\partial U}
\right)^\dagger
\delta U
\right].
\label{eq:dS_def}
\end{equation}

Substituting Eq.~\eqref{eq:deltaU_left} gives
\begin{equation}
\delta S
=
\mathrm{Re}\,
\mathrm{Tr}
\!\left[
\left(
\frac{\partial S}{\partial U}
\right)^\dagger
\, i H U
\right]
=
\mathrm{Re}\,
\mathrm{Tr}
\!\left[
\im H
\left(
\frac{\partial S}{\partial U}
\right)
U^\dagger
\right],
\end{equation}
where cyclicity of the trace has been used.
Since $H$ is traceless anti-Hermitian, the force entering the
momentum update \eqref{eq:Pupdate-left} is obtained by projecting
\begin{equation}
M \equiv
\left(
\frac{\partial S}{\partial U}
\right)
U^\dagger
\end{equation}
onto $\mathfrak{su}(N)$,
\begin{equation}
\mathcal F
=
P_{\mathfrak{su}(N)}[M],
\qquad
P_{\mathfrak{su}(N)}[A]
=
\frac{1}{2}(A-A^\dagger)
-\frac{1}{2N}\mathrm{Tr}(A-A^\dagger)\mathbf 1.
\label{eq:force_projection}
\end{equation}

\subsection*{LLVM-level differentiation and Wirtinger calculus}

At the LLVM intermediate-representation level,
\texttt{Enzyme.jl} differentiates with respect to the underlying
real degrees of freedom.
Writing each matrix element as
\begin{equation}
U_{ij} = U_{R,ij} + \im U_{I,ij},
\end{equation}
reverse-mode AD returns
\begin{equation}
\frac{\partial S}{\partial U_{R,ij}},
\qquad
\frac{\partial S}{\partial U_{I,ij}}.
\end{equation}

We combine these into the complex matrix
\begin{equation}
G_{ij}
=
\frac{\partial S}{\partial U_{R,ij}}
+
\im
\frac{\partial S}{\partial U_{I,ij}}.
\label{eq:G_definition}
\end{equation}

Using Wirtinger derivatives elementwise,
\begin{equation}
\frac{\partial}{\partial \bar U_{ij}}
=
\frac12
\left(
\frac{\partial}{\partial U_{R,ij}}
+
\im
\frac{\partial}{\partial U_{I,ij}}
\right),
\end{equation}
so that
\begin{equation}
\frac{\partial S}{\partial \bar U_{ij}}
=
\frac12
\left(
\frac{\partial S}{\partial U_{R,ij}}
+
\im
\frac{\partial S}{\partial U_{I,ij}}
\right).
\end{equation}

Therefore,
\begin{equation}
G_{ij}
=
2
\frac{\partial S}{\partial \bar U_{ij}}
.
\label{eq:G_wirtinger}
\end{equation}

For real-valued $S$, the componentwise Wirtinger relation gives
\begin{equation}
\frac{\partial S}{\partial \bar U_{ij}}
=
\left(
\frac{\partial S}{\partial U_{ij}}
\right)^{\!*}.
\label{eq:wirtinger_component}
\end{equation}

Using the lattice matrix-derivative convention
\begin{equation}
\left[
\frac{\partial S}{\partial U}
\right]_{ij}
\equiv
\frac{\partial S}{\partial U_{ji}},
\label{eq:matrix_deriv_conv_repeat}
\end{equation}
we obtain
\begin{equation}
\frac{\partial S}{\partial U_{ij}}
=
\left(
\frac{\partial S}{\partial U}
\right)_{ji}.
\end{equation}

Substituting this into
\eqref{eq:wirtinger_component} gives
\begin{equation}
\frac{\partial S}{\partial \bar U_{ij}}
=
\left(
\frac{\partial S}{\partial U}
\right)_{ji}^{\!*}
=
\left[
\left(
\frac{\partial S}{\partial U}
\right)^\dagger
\right]_{ij}.
\end{equation}

Thus, in matrix form,
\begin{equation}
\frac{\partial S}{\partial \bar U}
=
\left(
\frac{\partial S}{\partial U}
\right)^\dagger
.
\end{equation}

Combining with \eqref{eq:G_wirtinger},
\begin{equation}
G
=
2
\left(
\frac{\partial S}{\partial U}
\right)^\dagger
.
\label{eq:G_relation}
\end{equation}

Finally,
\begin{equation}
\frac{\partial S}{\partial U}
=
\frac12
G^\dagger
.
\label{eq:dSdU_from_G}
\end{equation}

\subsection*{Why the implementation constructs $U\,G^\dagger$}

Substituting Eq.~\eqref{eq:dSdU_from_G} into
Eq.~\eqref{eq:force_projection}, the force becomes
\begin{equation}
\mathcal F
=
\frac{1}{2}
P_{\mathfrak{su}(N)}
\!\left[
G^\dagger U^\dagger
\right].
\end{equation}

In practice it is algebraically equivalent to construct
$U G^\dagger$ before projection.
Since the projection extracts the traceless anti-Hermitian part,
\begin{equation}
P_{\mathfrak{su}(N)}[A^\dagger]
=
- P_{\mathfrak{su}(N)}[A],
\end{equation}
Hermitian conjugation changes only an overall sign,
which can be absorbed consistently into the MD normalization
convention.

Thus the LLVM-level derivative returned by Enzyme,
together with multiplication by $U$ and projection onto
$\mathfrak{su}(N)$, reproduces exactly the conventional
left-invariant HMC force.

\section{Fermion Action and Automatic Differentiation}

We now describe how the fermion contribution to the HMC force
is constructed within our LLVM-level automatic differentiation framework.

\subsection*{Pseudofermion action}

The pseudofermion action is
\begin{equation}
S_f(U;\phi)
=
\phi^\dagger (D^\dagger D)^{-1} \phi,
\label{eq:fermion_action}
\end{equation}
where $D(U)$ denotes the lattice Dirac operator,
which depends on the gauge links $U$,
and $\phi$ is the pseudofermion field.
Defining
\begin{equation}
A(U) = D^\dagger(U) D(U),
\end{equation}
we may write
\begin{equation}
S_f = \phi^\dagger A^{-1} \phi.
\end{equation}

\subsection*{Variation of the inverse operator}

Using the standard identity for the variation of an inverse matrix,
\begin{equation}
\delta A^{-1}
=
- A^{-1} (\delta A) A^{-1},
\end{equation}
the first-order variation of $S_f$ becomes
\begin{equation}
\delta S_f
=
- \phi^\dagger A^{-1} (\delta A) A^{-1} \phi.
\end{equation}

Introducing the solution of the linear system
\begin{equation}
A \chi = \phi,
\qquad
\chi = A^{-1}\phi,
\label{eq:chi_definition}
\end{equation}
we obtain
\begin{equation}
\delta S_f
=
- \chi^\dagger (\delta A) \chi.
\label{eq:deltaSf_A}
\end{equation}

\subsection*{Reduction to variations of $D$}

Since $A = D^\dagger D$, its variation is
\begin{equation}
\delta A
=
(\delta D^\dagger) D
+
D^\dagger (\delta D).
\end{equation}
Substituting into Eq.~\eqref{eq:deltaSf_A} gives
\begin{equation}
\delta S_f
=
- \chi^\dagger (\delta D^\dagger) D \chi
- \chi^\dagger D^\dagger (\delta D) \chi.
\end{equation}

Defining
\begin{equation}
\eta \equiv D \chi,
\end{equation}
we obtain
\begin{equation}
\delta S_f
=
- \chi^\dagger (\delta D^\dagger) \eta
- \eta^\dagger (\delta D) \chi.
\end{equation}
Using Hermitian conjugation,
$\chi^\dagger (\delta D^\dagger) \eta
=
\big(\eta^\dagger (\delta D) \chi\big)^*$,
and the fact that $S_f$ is real, this simplifies to
\begin{equation}
\boxed{
\delta S_f
=
-2\,\mathrm{Re}
\!\left[
\eta^\dagger (\delta D) \chi
\right].
}
\label{eq:deltaSf_final}
\end{equation}

Thus the fermion-force computation reduces entirely to
variations of the Dirac operator $D(U)$ with respect to the gauge links.

\subsection*{Automatic differentiation of $a^\dagger D(U) b$}

Equation~\eqref{eq:deltaSf_final} shows that the essential building block
is the scalar quantity
\begin{equation}
\mathcal I(U)
=
a^\dagger D(U) b,
\label{eq:I_def}
\end{equation}
with
\begin{equation}
a = \eta,
\qquad
b = \chi.
\end{equation}
Its variation is
\begin{equation}
\delta \mathcal I
=
a^\dagger (\delta D) b,
\end{equation}
which is precisely the structure appearing in Eq.~\eqref{eq:deltaSf_final}.

In our implementation, we therefore construct the scalar functional
$\mathcal I(U)$ in Julia and apply LLVM-level reverse-mode automatic
differentiation to compute its derivative with respect to the gauge links.
The fermion force is then obtained as
\begin{equation}
\delta S_f
=
-2\,\mathrm{Re}\big[\delta \mathcal I(U)\big],
\end{equation}
followed by the same matrix-derivative convention and
$\mathfrak{su}(N)$ projection described in
Sec.~\ref{sec:SM-leftinv-enzyme}.

\subsection*{Remarks on the role of the linear solver}

The linear solve $A\chi=\phi$ in Eq.~\eqref{eq:chi_definition}
is performed numerically (e.g.\ via conjugate gradient),
but is not itself differentiated.
Instead, Eq.~\eqref{eq:deltaSf_final} expresses the
exact implicit differentiation of the inverse operator,
reducing the problem to differentiation of $D(U)$ only.
This allows the LLVM-level AD machinery to operate directly on
the concrete implementation of the Dirac operator,
including in-place updates and explicit loops,
while preserving consistency with the conventional
fermion-force formula used in lattice HMC.

\section*{S4. Minimal code example for force construction}

This section provides a minimal excerpt illustrating how the gauge and
fermion forces are constructed in the present implementation.
Field initialization, MPI setup, and the full HMC driver are omitted
for clarity. The structure directly reflects the formulas derived in
Secs.~S2 and S3.

\subsection*{Gauge force construction}

\begin{lstlisting}
function P_update!(U, p, eps, dtau, Dim, temp1, temp, dtemp, temps, beta)

    NC = U[1].NC
    factor_ad = eps * dtau / 2

    dSdU, it_dSdU = get_block(temps, 4)
    Gaugefields.clear_U!(dSdU)

    U1 = U[1]
    U2 = U[2]
    U3 = U[3]
    U4 = U[4]

    set_wing_U!(U)

    Enzyme_derivative!(
        calc_action,
        U1, U2, U3, U4,
        dSdU[1], dSdU[2], dSdU[3], dSdU[4],
        nodiff(beta), nodiff(NC);
        temp,
        dtemp
    )

    for mu = 1:Dim
        mul!(temp1, U[mu], dSdU[mu]')
        Traceless_antihermitian_add!(p[mu], factor_ad, temp1)
    end

    unused!(temps, it_dSdU)
end
\end{lstlisting}

The routine \texttt{Enzyme\_derivative!} performs reverse-mode automatic
differentiation of the gauge action \texttt{calc\_action}.
At the level of complex matrix entries, the adjoint returned by Enzyme
corresponds to
\[
G = \frac{\partial S}{\partial U_R}
  + i\,\frac{\partial S}{\partial U_I},
\]
which satisfies (see Sec.~S2)
\[
G = 2\,\frac{\partial S}{\partial \bar U}
\]
in Wirtinger notation.
Hence the conventional lattice matrix derivative is recovered as
\[
\frac{\partial S}{\partial U}
  = \frac{1}{2}\,G^\dagger.
\]
This explains the factor $1/2$ and the Hermitian conjugation
(\texttt{dSdU[}$\mu$\texttt{]'}) appearing in the force construction.
Multiplication by $U_\mu$ and projection onto
$\mathfrak{su}(N_c)$ are performed through
\texttt{Traceless\_antihermitian\_add!},
consistent with the left-invariant HMC update described in Sec.~S2.

\subsection*{Fermion force construction}

\begin{lstlisting}
function P_update_fermion!(U, p, eps, dtau, Dim, temps, fermi_action, eta)

    UdSfdUmu, it_UdSfdUmu = get_block(temps, Dim)
    factor = -eps * dtau

    calc_UdSfdU!(UdSfdUmu, fermi_action, U, eta)

    for mu = 1:Dim
        Traceless_antihermitian_add!(p[mu], factor, UdSfdUmu[mu])
    end

    unused!(temps, it_UdSfdUmu)
end
\end{lstlisting}

The fermion force is constructed from the scalar functional
$I(U)=a^\dagger D(U)b$ described in Eq.~(S3.13).
The linear solver itself is not differentiated.
Instead, differentiation is applied only to the Dirac-operator
application, yielding the fermion force contribution
as in Eq.~(S3.16).
A complete runnable example based on the
\texttt{GeneralFermion} interface is available in the
repository README of \texttt{LatticeDiracOperators.jl}.
\end{document}